\documentclass[twoside,12pt]{article}
\usepackage{epsfig}

\usepackage{geometry}                		
\usepackage{graphicx}				
\usepackage{amssymb}
\usepackage{amsmath}
\usepackage{bm}
\usepackage{mathtools}
\usepackage{flexisym}
\usepackage{color}
\usepackage{float}

\begin{document}




\begin{center}
\LARGE\bf Slow-roll Analysis of Double Field Axion Inflation$^{*}$
\end{center}

\footnotetext{\hspace*{-.45cm}\footnotesize $^\dag$MAN Ping Kwan, Ellgan, E-mail:  ellgan101@akane.waseda.jp}

\begin{center}
\rm MAN Ping Kwan, Ellgan$^{\rm a)\dagger}$
\end{center}

\begin{center}
\begin{footnotesize} \sl
${Department \; of \; Pure \; and \; Applied \; Physics, \; Waseda \; University}^{\rm a)}$ \\
\end{footnotesize}
\end{center}


\vspace*{2mm}

\begin{center}
\begin{minipage}{15.5cm}
\parindent 20pt\footnotesize
We adopt the double field natural inflation model motivated by the non-perturbative effects of supergravity and superstring theory to do the slow roll analysis. We show that when the parameters are suitably chosen, there exist ranges of initial values of fields that can satisfy the constraints of Planck observations. This implies less fine tuning of field values can be allowed with tolerance of $4 {M}_{\text{pl}}$ for ${\phi}_{1}$ and $5 {M}_{\text{pl}}$ for ${\phi}_{2}$ respectively, which become more physical for field fluctuation in quantum era. We also show that the spectral index ${n}_{\mathcal{RR}}$, the fraction of entropic power spectrum and the fraction of power spectra (so-called ${\beta}_{\text{iso}}$) and $\cos{\Delta}$ can satisfy the constraints of Planck observation. This implies that double field natural inflation is a valid model to describe cosmological inflation. 
\end{minipage}
\end{center}

\begin{center}
\begin{minipage}{15.5cm}
\begin{minipage}[t]{2.3cm}{\bf Keywords:}\end{minipage}
\begin{minipage}[t]{13.1cm}
Superstring Theory, Double Field Inflation, Natural Inflation
\end{minipage}\par\vglue8pt

\end{minipage}
\end{center}

\section{Introduction}
\subsection{Present Observations for Inflation}
\noindent So far, there have been many theoretical approaches to study inflation dynamics, with the verification via Planck data. It has been shown that in single field inflation models, Starobinsky model\footnote{It is also called $R^2$ inflation model.}, power law model\footnote{It is only for $V \left( {\phi} \right) = {\lambda} {M}^{3}_{\text{pl}} {\phi}$.}, D-brane model\footnote{It is for $V \left( {\phi} \right) = {\Lambda} \left( 1 - \frac{{\mu}^{2}}{{\phi}^{p}}+ \cdots \right)$ where $p=2,4$. } and hilltop quartic model match with the Planck data between $50$ and $60$ e-folds \cite{Planck 2018 X}, while the constraints of slow-roll parameters, spectral index and its runnings are given in Table \ref{table:Planck data 2018 slow roll potential parameters and spectral indices}. 

\vspace{3mm}

\noindent Multiple scalar fields are usually included in inflation models under the motivation of high energy physics \cite{Particle physics models of inflation and the cosmological density perturbation} \cite{Particle physics models of inflation and curvaton scenarios}. Unlike single field models, in which only adiabatic perturbation is present, multifield models generically produce entropic perturbation apart from adiabatic counterpart. The entropic perturbation can contribute to the adiabatic perturbation, thereby supporting inflation dynamics. \cite{Cosmological Perturbation Theory Part 2} \cite{Primordial Bispectrum from Multi-field Inflation with Non-minimal Couplings}. Hence, understanding the evolution of entropic perturbation and its coupling with adiabatic one are vital for studying non-trivial features of inflation that are not present in the single field cases. 

\vspace{3mm}

\noindent Even though single field natural inflation can be derived from superstring theory \cite{Natural Inflation and Low Energy Supersymmetry}, recent investigation shows that it is disfavoured by Planck 2018 observation \cite{Planck 2018 X}. The reasons are because it has a low Bayes factor of $\ln{B}=-4.2$ and its prediction of the tensor-to-scalar ratio against spectral index cannot enter the innermost region of Planck 2018 observation \cite{Planck 2018 X}. 

\begin{table}[h]
\begin{center}
\begin{tabular}{ |c|c|c|c| }
\hline
$\text{Slow-roll parameters}$ & $\text{Range(s)}$ & $\text{Spectral indices}$ & $\text{Range(s)}$ \\
\hline
${\epsilon}_{V}$ & $<0.0053$ & ${n}_{s} - 1$ & $[-0.0423, -0.0327]$ \\
\hline
${\eta}_{V}$ & $[-0.021, -0.008]$ & ${\alpha}_{s} := \frac{d {n}_{s}}{d \ln{k}}$ & $[-0.008, 0.012]$ \\
\hline
${\xi}_{V}$ & $[-0.0045, 0.0096]$ & ${\beta}_{s} := \frac{d^{2} {n}_{s}}{d \ln{k}^{2}}$ & $[-0.003, 0.023]$ \\
\hline
$$ & $$ & $\ln(10^{10} A_{s})$ & $[3.03, 3.058]$ \\
\hline
\end{tabular}
\end{center}
\caption{Slow roll potential parameters and spectral indices in Planck 2018}
\label{table:Planck data 2018 slow roll potential parameters and spectral indices}
\end{table}

\subsection{Main theme of this paper}
\noindent Despite this, nowadays, superstring theory is the most promising theory for quantum gravity. Some considered the possibility of multi-field inflation \cite{Cosmological perturbations with multiple scalar fields}. In particular, some investigated the possibility of double inflation dynamics \cite{Curvature and iso-curvature perturbations from two-field inflation in a slow-roll expansion}. Thus, we raise questions, "What will be the dynamics if double field natural inflation is considered? Can it satisfy the recent Planck observational constraints?" In this paper, we show that when the parameters are suitably chosen, there exist ranges of initial values of fields that can satisfy the constraints of Planck observations. Particularly, the predictions of spectral index ${n}_{\mathcal{RR}}$, tensor-to-scalar ratio $r$, ${\beta}_{\text{iso}}$ and $\cos{\Delta}$ are within the ranges of the present observation. This means that some specially chosen parameters can allow less fine tuning of field values, which become more physical for field fluctuation in quantum era\footnote{Note that the beginning of inflation comes from the end of quantum era.}. \\

\vspace{3mm}

\noindent The arrangement of this paper is as follows.\footnote{In this paper, we set the reduced Planck mass as ${M}_{\text{pl}}$.} In section \ref{Deviation}, starting from a short review of single field natural inflation, we show how the double field natural inflation can be obtained by field stabilisation, whose ideas is to set those fields irresponsible for the inflation as a certain value such that the resulting inflation potential can be minimum along that field direction. In section \ref{Formalism}, following the arguments in \cite{Primordial Bispectrum from Multi-field Inflation with Non-minimal Couplings} and \cite{Multifield inflation after Planck: iso-curvature modes from non-minimal couplings}, we introduce the concepts of entropic perturbation, and derive the corresponding spectral index ${n}_{\mathcal{RR}}$, tensor-to-scalar ratio $r$, ${\beta}_{\text{iso}}$ and $\cos{\Delta}$. In section \ref{Numerical Calculations and Results}, we show how less fine tuning can be achieved by some choices of parameters, and evaluate all the related inflation parameters accordingly. Finally, in section \ref{Discussion}, we explain the data. \\

\section{Mathematical Derivation}
\label{Deviation}
\subsection{A short review of single field natural inflation}
\noindent Axions are hypothetical particles associated with the spontaneously broken Peccei-Quinn (PQ) symmetry that can solve the strong CP problem in QCD \cite{CP conservation in the presence of pseudoparticles} \cite{Constraints imposed by CP conservation in the presence of pseudoparticles}. After that, physicists adopted the idea of axion in superstring theory that they realize aligned natural inflation on the type IIB orientifold compactification with fluxes \cite{Natural Inflation and Low Energy Supersymmetry}. Due to the breaking from continuous symmetry\footnote{It means that for all complex scalars $c$, the transformation ${\phi} \rightarrow {\phi} + c$ makes the Lagrangian invariant. } into discrete shift symmetry\footnote{\noindent It means that unless the transformation is in the form ${\phi} \rightarrow {\phi} + 2 \pi f$ , where $f$ is the corresponding axion decay constant, the Lagrangian is no longer invariant under shift transformations other than this form. } by non perturbative effects such as instanton effect and gaugino condensation, the natural inflation potential is produced as a sinusoidal function. For single field case, we have the potential

\begin{equation}
\label{Single Field Natural Inflation}
V({\phi}) = {\Lambda}^{4} \left[ 1 - \cos{\left( \frac{\phi}{f} \right)} \right], \\
\end{equation}
\noindent where ${\Lambda}$ and $f$ are inflation amplitude and axion decay constant respectively, with the scale $f \geq 10 {M}_{\text{pl}}$ and ${\Lambda} \approx 10^{-3} M_{\text{pl}}$ \cite{Aligned Natural Inflation in String Theory}. This model can be motivated from various directions like supergravity embeddings \cite{Natural inflation in supergravity and beyond}. Later on, models whose axion symmetries are broken by more than one instanton were proposed. \\ 

\subsection{Field Stabilisation} 
\noindent The modulated natural inflation model can be motivated by adopting the F term potential of supergravity \cite{Supergravity}. \footnote{There are 2 terms in supergravity potential $V = V_{F} + V_{D}$. But now that the F term potential can produce a non-negative energy environment, we can neglect the D term for simplicity. However, in general, if the F term cannot produce a non-negative energy environment, D term is required to correct the negative F term. For reference, please see the KKLT model \cite{de Sitter Vacua in String Theory}. }

\begin{equation} 
\label{{V}_{F}}
{V}_{F} = e^{\frac{K}{{M}^{2}_{\text{pl}}}} \left[ K^{{\alpha}{\bar{\beta}}} {D}_{\alpha}W \overline{{D}_{{\beta}} {W}} - \frac{3 |W|^{2}}{{M}^{2}_{\text{pl}}} \right]= e^{K} \left[ {K}^{e \bar{f}} D_{e}W \overline{D_{f} W} - 3 |W|^{2} \right], \\
\end{equation}

\noindent where $W$ and $K$ are super-potential and K\"{a}hler potential respectively, $K^{e \bar{f}}$ is the inverse of the Hessian matrix of the K\"{a}hler potential with respect to scalar fields ${\phi}_{e}$ and $\bar{\phi}_{f}$, and the over bar means the conjugate. For the rest of the paper, we take ${M}_{\text{pl}} = 1$. Now, we consider the following super-potential $W({\rho}_{1}, {\rho}_{2}, {X}_{1}, {X}_{2})$ and K\"{a}hler potential $K({\rho}_{1}, {\rho}_{2}, {X}_{1}, {X}_{2})$ \cite{Natural Inflation and Low Energy Supersymmetry}

\begin{equation}
W({\rho}_{1}, {\rho}_{2}, {X}_{1}, {X}_{2}) = \sum^{2}_{j=1} m^2_{j} X_{j} \left( e^{-a_{j}{\rho}_{1} - b_{j} {\rho}_{2}} - {\lambda}_{j} \right), \\
\end{equation}
\begin{equation}
K({\rho}_{1}, {\rho}_{2}, {X}_{1}, {X}_{2}) =  - \sum^{2}_{j=1} \left[ \ln{\left( {\rho}_{j} + \bar{\rho}_{j} \right)} - k |X_{j}|^{2} \right], \\
\end{equation}

\noindent where $X_{1,2}$ and ${\rho}_{1, 2}$ are matter fields and moduli respectively\footnote{Note that one modulus field ${\rho}_{j}$ consists of saxion field ${\chi}_{j}$ and axion field $\tilde{{\phi}}_{j}$ with ${\chi}_{j}, \tilde{{\phi}}_{j} \in \mathbb{R} \; \forall j \in \{1,2 \}$. That is ${\rho}_{j} = {\chi}_{j} + i \tilde{{\phi}}_{j} $. }. The $m_{1,2}$\footnote{${m}_{j} \in \mathbb{R}$. } and ${\lambda}_{1,2}$\footnote{The modulus ${\lambda}_{j}$ is defined as ${\lambda}_{j} = |{\lambda}_{j}| e^{i {\alpha}_{j}}$ with $|{\lambda}_{j}| \in \mathbb{R}, \; {\alpha}_{j} \in [ 0, 2 \pi) \; \forall j \in \{ 1,2 \}$. } are other chiral matter fields and moduli, which will stabilise the the potential with non-zero vacuum expectation values (VEVs) \cite{Natural Inflation and Low Energy Supersymmetry}. Since the stabilisation scale is high enough, they can be treated as constants \cite{Natural Inflation and Low Energy Supersymmetry}. The exponential terms in the super-potential come from the D brane instantons and/or gaugino condensates \cite{Flux Compactifications in String Theory a Comprehensive Review}. \\

\noindent To start the field stabilisation, we first find the SUSY ground state. This is equivalent to the Supersymmetry (SUSY) preservation $D_{i}W = 0$. By using the definition $D_{i}W = {\partial}_{i} W + \left( {\partial}_{i} K \right) W = 0$ for all fields $X_{1}, X_{2}, {\rho}_{1}, {\rho}_{2}$, we obtain 
\begin{equation}
\begin{split}
{X}_{1} =& \; {X}_{2} = 0, \\
{\chi}_{1} =& \; \frac{ b_{2} \ln{|{\lambda}_{1}|} - b_{1} \ln{|{\lambda}_{2}|} }{ a_{2} b_{1} - a_{1} b_{2} } \equiv {\chi}_{1,0}, \\
{\chi}_{2} =& \; - \frac{ a_{2} \ln{|{\lambda}_{1}|} - a_{1} \ln{|{\lambda}_{2}|} }{ a_{2} b_{1} - a_{1} b_{2} } \equiv {\chi}_{2,0}, \\
\end{split}
\end{equation}
\noindent and 
\begin{equation}
\begin{split}
{a}_{1} \tilde{{\phi}}_{1,0} + {b}_{1} \tilde{{\phi}}_{2,0} + {\alpha}_{1} \in& \; 2 \pi \mathbb{Z}, \\
{a}_{2} \tilde{{\phi}}_{1,0} + {b}_{2} \tilde{{\phi}}_{2,0} + {\alpha}_{2} \in& \; 2 \pi \mathbb{Z}. \\
\end{split}
\end{equation}
\noindent During inflation, it is assumed that only the axion fields run while the saxion fields and the stabilisers are kept in their minimum. Thus, by taking ${X}_{j} = 0$ and ${\rho}_{j} = {\chi}_{j, 0} + i \tilde{{\phi}}_{j}, \; \forall j \in \{1,2\}$ and using the F-term potential Eq.(\ref{{V}_{F}}), we obtain
\begin{equation}
{V}(\tilde{{\phi}}_{1}, \tilde{{\phi}}_{2}) = \frac{1}{ 2 k {\chi}_{1,0} {\chi}_{2,0} } \sum^{2}_{j=1} \left \{ m^{4}_{j} |{\lambda}_{j}|^{2} \left[ 1 - \cos{({a}_{j} \tilde{{\phi}}_{1} + {b}_{j} \tilde{{\phi}}_{2})} \right] \right \}. \\
\end{equation}

\noindent Hence, we get the double field natural inflation model from this motivation. 

\section{Formalism}
\label{Formalism}
\noindent In this section, we follow the derivation in \cite{Primordial Bispectrum from Multi-field Inflation with Non-minimal Couplings} and \cite{Multifield inflation after Planck: iso-curvature modes from non-minimal couplings}. Note that in the Jordan frame, the Lagrangian is
\begin{equation}
{S}_{\text{Jordan}} = \int d^{4} x \sqrt{-\tilde{g}} \left[ f\left({\phi}^{I}\right) \tilde{R} - \frac{1}{2} \tilde{\mathcal{G}}_{IJ} \tilde{g}^{{\mu}{\nu}} \tilde{\triangledown}_{\mu} {\phi}^{I} \tilde{\triangledown}_{\nu} {\phi}^{J} - \tilde{V}\left( {\phi}^{I} \right) \right]. \\
\end{equation}
\noindent where $f\left( {\phi}^{I} \right)$ is the non-minimal coupling function and $\tilde{V}\left( {\phi}^{I} \right)$ is the potential for the scalar fields in the Jordan frame. To change the equation in Jordan frame into the counterpart in Einstein frame, we define a spacetime metric in the Einstein frame ${g}_{{\mu}{\nu}}\left( x \right)$ as 
\begin{equation}
{g}_{{\mu}{\nu}}\left( x \right) = {\Omega}^{2} \left( x \right) \tilde{g}_{{\mu}{\nu}}\left( x \right), \\
\end{equation}
\noindent where the conformal factor ${\Omega}^{2} \left( x \right)$ is given by
\begin{equation}
{\Omega}^{2} \left( x \right) = \frac{2}{{M}^{2}_{\text{pl}}} f \left( {\phi}^{I} \left( x \right) \right). \\
\end{equation}
\noindent Then, the action in Jordan frame becomes that in Einstein frame, which is given by
\begin{equation}
{S}_{\text{Einstein}} = \int d^{4} x \sqrt{-{g}} \left[ \frac{{M}^{2}_{\text{pl}}}{2} {R} - \frac{1}{2} {\mathcal{G}}_{IJ} {g}^{{\mu}{\nu}} {\triangledown}_{\mu} {\phi}^{I} {\triangledown}_{\nu} {\phi}^{J} - {V}\left( {\phi}^{I} \right) \right]. \\
\end{equation}
\noindent and the potential in the Einstein frame becomes
\begin{equation}
\label{Action in Einstein frame}
V \left( {\phi}^{I} \right) = \frac{\tilde{V} \left( {\phi}^{I} \right)}{{\Omega}^{4} \left( x \right)} = \frac{{M}^{4}_{\text{pl}}}{4 f^2 \left( {\phi}^{I} \right)} \tilde{V} \left( {\phi}^{I} \right). \\
\end{equation}
\noindent The coefficients $\mathcal{G}_{{I}{J}}$ of the non-canonical kinetic terms in the Einstein frame depend on the non-minimal coupling function $f\left( {\phi}^{I} \right)$ and its derivatives. They are given by
\begin{equation}
\mathcal{G}_{{I}{J}} \left( {\phi}^{K} \right) = \frac{{M}^{2}_{\text{pl}}}{2 f \left( {\phi}^{L} \right)} \left[ \tilde{\mathcal{G}}_{{I}{J}} \left( {\phi}^{K} \right) + \frac{3}{f \left( {\phi}^{L} \right)} {f}_{,I} {f}_{,J} \right], \\
\end{equation}
\noindent where ${f}_{,I}=\frac{{\partial}{f}}{{\partial}{\phi}^{I}}$. Varying the action in Einstein frame with respect to ${g}_{{\mu}{\nu}} \left( x \right)$, we have the Einstein equations
\begin{equation}
{R}_{{\mu}{\nu}} - \frac{1}{2}{g}_{{\mu}{\nu}} R = \frac{1}{{M}^{2}_{\text{pl}}} {T}_{{\mu}{\nu}}, \\
\end{equation}
\noindent where 
\begin{equation}
{T}_{{\mu}{\nu}} = \mathcal{G}_{{I}{J}} {\partial}_{\mu} {\phi}^{I} {\partial}_{\nu} {\phi}^{J} - {g}_{{\mu}{\nu}} \left[ \frac{1}{2} \mathcal{G}_{{K}{L}} {\partial}_{\alpha} {\phi}^{K} {\partial}^{\alpha} {\phi}^{L} + V\left( {\phi}^{K} \right) \right]. \\
\end{equation}
\noindent Varying Eq. (\ref{Action in Einstein frame}) with respect to ${\phi}^{I}$, we obtain the equation of motion for ${\phi}^{I}$
\begin{equation}
\square {\phi}^{I} + {g}^{{\mu}{\nu}} {\Gamma}^{I}_{{J}{K}} {\partial}_{\mu} {\phi}^{J} {\partial}_{\nu} {\phi}^{K} - \mathcal{G}^{{I}{K}} {V}_{,K} = 0, \\ 
\end{equation}
\noindent where $\square {\phi}^{I} = {g}^{{\mu}{\nu}} {\phi}^{I}_{;{\mu}{\nu}}$ and ${\Gamma}^{I}_{{J}{K}}$ is the Christoffel symbol for the field space manifold in terms of $\mathcal{G}_{{I}{J}}$ and its derivative. Expanding each scalar field to the first order around its classical background value, 
\begin{equation}
{\phi}^{I} \left( {x}^{\mu} \right) = {\varphi}^{I} \left( t \right) + {\delta} {\phi}^{I} \left( {x}^{\mu} \right), \\
\end{equation}
\noindent and perturbing a spatially flat Friedmann-Robertson-Walker (FRW) metric,
\begin{equation}
ds^2={g}_{{\mu}{\nu}} d{x}^{\mu} d{x}^{\nu} = - \left(1+2A \right) dt^2 + 2 a \left( {\partial}_{i} B \right) d{x}^{i} dt + a^2 \left[ \left( 1-2{\psi} \right) {\delta}_{ij} + 2 {\partial}_{i} {\partial}_{j} E \right] d{x}^{i} d{x}^{j}, \\
\end{equation}
\noindent where $a\left( t \right)$ is the scale factor. To the zeroth order, the $00$ and $ij$ components of the Einstein equations become
\begin{equation}
H^2 = \frac{1}{3 {M}^{2}_{\text{pl}}} \left[ \frac{1}{2} \mathcal{G}_{{I}{J}} \dot{\varphi}^{I} \dot{\varphi}^{J} + V \left( {\varphi}^{I} \right) \right], \\
\end{equation}
\begin{equation}
\dot{H} = - \frac{1}{2 {M}^{2}_{\text{pl}}} \mathcal{G}_{{I}{J}} \dot{\varphi}^{I} \dot{\varphi}^{J}, \\
\end{equation}
\noindent where $H = \frac{\dot{a} \left( t \right)}{a \left( t \right)}$ is the Hubble parameter, and the field field space metric is calculated at the zeroth order, $\mathcal{G}_{{I}{J}} = \mathcal{G}_{{I}{J}} \left( {\varphi}^{K} \right)$. Introducing the number of e-folding $N=\ln{a}$ with $d N = H dt$, the above Einstein equation becomes
\begin{equation}
3 {M}^{2}_{\text{pl}} - \frac{1}{2} \mathcal{G}_{{I}{J}} {{\varphi}^{I}}' {{\varphi}^{J}}' = \frac{V \left( {\varphi}^{I} \right)}{H^2}, \\
\end{equation}
\begin{equation}
\frac{H'}{H} = - \frac{1}{2 {M}^{2}_{\text{pl}}} \mathcal{G}_{{I}{J}} {{\varphi}^{I}}' {{\varphi}^{J}}', \\
\end{equation}
\noindent where the prime $'$ means the derivative with respect to $N$. For any vector in the field space $A^{I}$, we define a covariant derivative with respect to the field-space metric as usual by
\begin{equation}
\mathcal{D}_{J} {A}^{I} = {\partial}_{J} {A}^{I} + {\Gamma}^{I}_{{J}{K}} {A}^{K}, \\
\end{equation}
\noindent and the time derivative with respect to the cosmic time $t$ is given by
\begin{equation}
\mathcal{D}_{t} {A}^{I} \equiv \dot{\varphi}^{J} \mathcal{D}_{J} {A}^{I} = \dot{A}^{I} + {\Gamma}^{I}_{{J}{K}} \dot{\varphi}^{J} {A}^{K} = H \left( {{A}^{I}}' + {\Gamma}^{I}_{{J}{K}} {{\varphi}^{J}}' {A}^{K} \right). \\
\end{equation}
\noindent Now, we define the length of the velocity vector for the background fields as
\begin{equation}
|\dot{\varphi}^{I}| \equiv \dot{\sigma} = \sqrt{\mathcal{G}_{{P}{Q}} \dot{\varphi}^{P} \dot{\varphi}^{Q}} \quad \Rightarrow \quad |{{\varphi}^{I}}'| \equiv {\sigma}' = \sqrt{\mathcal{G}_{{P}{Q}} {{\varphi}^{P}}' {{\varphi}^{Q}}'}. 
\end{equation}
\noindent Introducing the unit vector of the velocity vector of the background fields
\begin{equation}
\hat{\sigma}^{I} \equiv \frac{\dot{\varphi}^{I}}{\dot{\sigma}} = \frac{{{\varphi}^{I}}'}{{\sigma}'} = \frac{{{\varphi}^{I}}'}{\sqrt{\mathcal{G}_{{P}{Q}} {{\varphi}^{P}}' {{\varphi}^{Q}}'}} 
\end{equation}
\noindent the $00$ and $ij$ components of the Einstein equations become
\begin{equation}
H^2 = \frac{1}{3 {M}^{2}_{\text{pl}}} \left[ \frac{1}{2} \dot{\sigma}^2 + V \right] \quad \Rightarrow \quad 3 {M}^{2}_{\text{pl}} - \frac{1}{2} {{\sigma}'}^{2} = \frac{V \left( {\varphi}^{I} \right)}{H^2}, \\
\end{equation}
\begin{equation}
\dot{H} = - \frac{1}{2 {M}^{2}_{\text{pl}}} \dot{\sigma}^2 \quad \Rightarrow \quad \frac{H'}{H} = - \frac{1}{2 {M}^{2}_{\text{pl}}} {{\sigma}'}^{2}, \\
\end{equation}
\noindent and the equation of motion of ${\phi}^{I}$ in the zeroth order is 
\begin{equation}
\ddot{\sigma} + 3 H \dot{\sigma} + {V}_{,\sigma} = 0, \\
\end{equation}
\noindent where
\begin{equation}
{V}_{,\sigma} \equiv \hat{\sigma}^{I} {V}_{,I}. \\
\end{equation}
\noindent Now, we define a quantity $\hat{s}^{{I}{J}}$ to obtain the field component orthogonal to $\hat{\sigma}^{I}$
\begin{equation}
\hat{s}^{{I}{J}} \equiv \mathcal{G}^{{I}{J}} - \hat{\sigma}^{I} \hat{\sigma}^{J}, \\
\end{equation}
\noindent which obeys the following relations with $\hat{\sigma}^{I}$
\begin{equation}
\begin{split}
\hat{\sigma}_{I} \hat{\sigma}^{I} = 1, \\
\hat{s}^{{I}{J}} \hat{s}_{{I}{J}} = \mathcal{N} - 1, \\
\hat{s}^{I}_{\; A} \hat{s}^{A}_{\; J} = \hat{s}^{I}_{\; J}, \\
\hat{\sigma}_{I} \hat{s}^{{I}{J}} = 0 \quad \forall J. \\
\end{split}
\end{equation}
\noindent The slow roll parameters are given by
\begin{equation}
\epsilon \equiv - \frac{\dot{H}}{H^2} = \frac{3 \dot{\sigma}^2}{\dot{\sigma}^2 + 2 V}, \\
\end{equation}
\noindent and
\begin{equation}
{\eta}_{{\sigma}{\sigma}} \equiv {M}^{2}_{\text{pl}} \frac{\mathcal{M}_{{\sigma}{\sigma}}}{V} \quad \text{and} \quad {\eta}_{{s}{s}} \equiv {M}^{2}_{\text{pl}} \frac{\mathcal{M}_{{s}{s}}}{V}, \\
\end{equation}
\noindent where 
\begin{equation}
\begin{split}
{M}_{{\sigma}{I}} \equiv& \; \hat{\sigma}_{I} {M}^{I}_{J} = \hat{\sigma}^{K} \left( \mathcal{D}_{K} \mathcal{D}_{J} V \right), \\
{M}_{{\sigma}{\sigma}} \equiv& \; \hat{\sigma}_{I} \hat{\sigma}^{J} {M}^{I}_{J} = \hat{\sigma}^{K} \hat{\sigma}^{J} \left( \mathcal{D}_{K} \mathcal{D}_{J} V \right), \\
{M}_{{s}{s}} \equiv& \; \hat{s}_{I} \hat{s}^{J} {M}^{I}_{J} = \hat{s}^{K} \hat{s}^{J} \left( \mathcal{D}_{K} \mathcal{D}_{J} V \right). \\
\end{split}
\end{equation}
\noindent and $\hat{s}^{I}$ is defined in the following argument. Now we define the turn-rate vector ${\omega}^{I}$ as the covariant rate of change of the unit vector $\hat{\sigma}^{I}$
\begin{equation}
{\omega}^{I} \equiv \mathcal{D}_{t} \hat{\sigma}^{I} = - \frac{1}{\dot{\sigma}} {V}_{,K} \hat{s}^{{I}{K}}. \\
\end{equation}
\noindent Since ${\omega}^{I} \propto \hat{s}^{{I}{K}}$, we have
\begin{equation}
{\omega}^{I} \hat{\sigma}_{I} = 0. \\
\end{equation}
\noindent We can also find 
\begin{equation}
\mathcal{D}_{t} \hat{s}^{{I}{J}} = - \hat{\sigma}^{I} {\omega}^{J} - \hat{\sigma}^{J} {\omega}^{I}. \\
\end{equation}
\noindent Also, we introduce a new unit vector $\hat{s}^{I}$ pointing in the direction of the turn-rate, ${\omega}^{I}$, and a new projection operator ${\gamma}^{{I}{J}}$
\begin{equation}
\hat{s}^{I} \equiv \frac{{\omega}^{I}}{\omega}, \\
\end{equation}
\begin{equation}
{\gamma}^{{I}{J}} \equiv {\mathcal{G}}^{{I}{J}} - \hat{\sigma}^{I} \hat{\sigma}^{J} - \hat{s}^{I} \hat{s}^{J}. \\
\end{equation}
\noindent where ${\omega} = |{\omega}^{I}|$ is the magnitude of the turn-rate vector. The new unit vector $\hat{s}^{I}$ and the new projection operator ${\gamma}^{{I}{J}}$ also satisfy
\begin{equation}
\begin{split}
\hat{s}^{{I}{J}} =& \hat{s}^{I} \hat{s}^{J} + {\gamma}^{{I}{J}}, \\
{\gamma}^{{I}{J}} {\gamma}_{{I}{J}} =& \mathcal{N} -2, \\
\hat{s}^{{I}{J}} \hat{s}_{J} =& \hat{s}^{I}, \\
\hat{\sigma}_{I} \hat{s}^{I} =& \hat{\sigma}_{I} {\gamma}^{{I}{J}} = \hat{s}_{I} {\gamma}^{{I}{J}} = 0 \quad \forall J.
\end{split}
\end{equation}
\noindent We then find 
\begin{equation}
{\mathcal{D}}_{t} \hat{s}^{I} = - {\omega} \hat{\sigma}^{I} - {\Pi}^{I}, \\
{\mathcal{D}}_{t} {\gamma}^{{I}{J}} = \hat{s}^{I} {\gamma}^{J} + \hat{s}^{J} {\gamma}^{I}, \\
\end{equation}
\noindent where
\begin{equation}
{\Pi}^{I} \equiv \frac{1}{\omega} {\mathcal{M}}_{{\sigma}{K}} {\gamma}^{{I}{K}}, \\
\end{equation}
\noindent and hence
\begin{equation}
\hat{\sigma}_{I} {\Pi}^{I} = \hat{s}_{I} {\Pi}^{I} = 0, \\
\end{equation}
\noindent Now, we define the curvature and entropic perturbations as follows
\begin{equation}
\mathcal{R} = {\psi} + \frac{H}{\dot{\sigma}} \hat{\sigma}_{J} {\delta} {\phi}^{J} = \frac{H}{\dot{\sigma}} {Q}_{\sigma}, \\
\end{equation}
\begin{equation}
\mathcal{S} = \frac{H}{\dot{\sigma}} {Q}_{s}. \\
\end{equation}
\noindent After the first horizon crossing, the co-moving wave number $k$ obeys $\frac{k}{aH}<1$. Hence, the curvature and entropic perturbations satisfy the following equations
\begin{equation}
\dot{\mathcal{R}} = {\alpha} H \mathcal{S} + O \left( \frac{k^2}{a^2 H^2} \right), \\
\end{equation}
\begin{equation}
\dot{\mathcal{S}} = {\beta} H \mathcal{S} + O \left( \frac{k^2}{a^2 H^2} \right). \\
\end{equation}
\noindent In the double field case, we assume the perturbation spectra evolve outside the horizon exit. This is different from the single field case, where the curvature power spectrum remains unchanged outside the horizon \cite{Cosmological Perturbation Theory Part 2}. The curvature and entropy perturbations at some cosmic time $t$ are assumed to be proportional to the corresponding values at the horizon exit, which are given by the following matrix transformation \cite{Cosmological Perturbation Theory Part 2}.

\begin{equation}
\left(
\begin{array}{rr}
\mathcal{R} (t) \\
\mathcal{S} (t) \\
\end{array}
\right) = \left(
\begin{array}{rr}
T_{\mathcal{RR}} (t, {t}_{\text{hc}}) & T_{\mathcal{RS}} (t, {t}_{\text{hc}}) \\
T_{\mathcal{SR}} (t, {t}_{\text{hc}}) & T_{\mathcal{SS}} (t, {t}_{\text{hc}}) \\
\end{array}
\right) \left(
\begin{array}{rr}
\mathcal{R} ({t}_{\text{hc}}) \\
\mathcal{S} ({t}_{\text{hc}}) \\
\end{array}
\right), \\
\end{equation}
\noindent where $T_{\mathcal{RS}} (t, {t}_{\text{hc}})$\footnote{For those who can be easy to remember, note that the subscript of $T_{\mathcal{RS}}$ and the time flow should be read from right to left.} means the transfer function from entropic perturbation to curvature perturbation from the time at the horizon exit to some later cosmic time $t$. In general, we assume curvature perturbation does not evolve to the entropic counterpart, $T_{\mathcal{SR}} (t, {t}_{\text{hc}}) = 0$, and the transition function from curvature perturbation to itself remains constant, $T_{\mathcal{RR}} (t, {t}_{\text{hc}}) = 1$ \cite{Cosmological Perturbation Theory Part 2}. Thus, the above matrix transformation becomes
\begin{equation} \label{Perturbation Equation}
\left(
\begin{array}{rr}
\mathcal{R} (t) \\
\mathcal{S} (t) \\
\end{array}
\right) = \left(
\begin{array}{rr}
1 & T_{\mathcal{RS}} (t, {t}_{\text{hc}}) \\
0 & T_{\mathcal{SS}} (t, {t}_{\text{hc}}) \\
\end{array}
\right) \left(
\begin{array}{rr}
\mathcal{R} ({t}_{\text{hc}}) \\
\mathcal{S} ({t}_{\text{hc}}) \\
\end{array}
\right). \\
\end{equation}
\noindent Now, the transfer functions are given by
\begin{equation}
{T}_{\mathcal{RS}} \left( {t}_{\text{hc}}, t \right) = \int^{t}_{{t}_{\text{hc}}} d {t}' {\alpha} \left( t' \right) H \left( t' \right) {T}_{\mathcal{SS}} \left( {t}_{\text{hc}}, t' \right), \\
\end{equation}
\begin{equation}
{T}_{\mathcal{SS}} \left( {t}_{\text{hc}}, t \right) = \exp \left[ \int^{t}_{{t}_{\text{hc}}} {dt'}{\beta} \left( t' \right) H \left( t' \right) \right]. \\
\end{equation}
\noindent where ${t}_{\text{hc}}$ is the time of the first horizon crossing and ${k}^2_{*} = {a}^{2} \left( {t}_{\text{hc}} \right) {H}^{2} \left( {t}_{\text{hc}} \right)$. Being changed from the cosmic time $t$ into the number of e-folding $N=\ln{a}$, where $dN=Hdt$\footnote{In some literatures like \cite{Multifield inflation after Planck: iso-curvature modes from non-minimal couplings}, ${N}_{*} = {N}_{\text{tot}} - N\left( t \right)$ is used and the corresponding differential equation becomes $d {N}_{*} = - H dt$. But, in this paper, we keep using $dN=Hdt$.}, ${T}_{\mathcal{RS}} \left( {t}_{\text{hc}}, t \right)$ and ${T}_{\mathcal{SS}} \left( {t}_{\text{hc}}, t \right)$ become 
\begin{equation}
{T}_{\mathcal{RS}} \left( {N}_{\text{hc}}, N \right) = \int^{N}_{{N}_{\text{hc}}} {dN'} {\alpha} \left( N' \right) {T}_{\mathcal{SS}} \left( {N}_{\text{hc}}, N' \right), \\
\end{equation}
\noindent and
\begin{equation}
{T}_{\mathcal{SS}} \left( {N}_{\text{hc}}, N \right) = \exp \left[ \int^{N}_{{N}_{\text{hc}}} {dN'} {\beta} \left( N' \right) \right]. \\
\end{equation}
\noindent Now that \cite{Primordial Bispectrum from Multi-field Inflation with Non-minimal Couplings}
\begin{equation}
\dot{\mathcal{R}} = 2 {\omega} \mathcal{S} + O \left( \frac{k^2}{a^2 H^2} \right), \\
\end{equation}
\noindent and
\begin{equation}
\dot{Q}_{s} \simeq - \frac{{\mu}^{2}_{s}}{3 H} {Q}_{s}, \\
\end{equation}
\noindent where ${\mu}^{2}_{s} = {\mathcal{M}}_{\text{ss}} + 3 {\omega}^{2}$ and $\simeq$ means slow-roll approximation with an aid of an equation
\begin{equation}
3 H \dot{\sigma} \simeq - \hat{\sigma}^{I} {V}_{,I}. \\
\end{equation}
\noindent Then, we find
\begin{equation}
{\alpha} \left( t \right) = \frac{2 {\omega} \left( t \right)}{H \left( t \right)}, \\
\end{equation}
\noindent and 
\begin{equation}
{\beta} = - \frac{{\mu}^{2}_{s}}{3 H^2} - {\epsilon} - \frac{\ddot{\sigma}}{H \dot{\sigma}} = -\frac{1}{4} {\eta}_{ss} \left( 1 - \frac{1}{3}{\epsilon} \right) + \left( 3 - {\epsilon} \right) + \frac{3 - {\epsilon}}{2 {\epsilon}} \frac{d}{d N} \left({\ln{V}} \right) - \frac{{\omega}^{2}}{H^2}, \\
\end{equation}
\noindent Note that the power spectrum for the gauge invariant curvature perturbation is given by
\begin{equation}
\langle \mathcal{R}_{c} \left( \bold{k}_{1} \right) \mathcal{R}_{c} \left( \bold{k}_{2} \right) \rangle = \left( 2 \pi \right)^{3} {\delta}^{\left( 3 \right)} \left( \bold{k}_{1} + \bold{k}_{2} \right) {P}_{\mathcal{R}} \left( {k}_{1} \right), \\
\end{equation}
\noindent where ${P}_{\mathcal{R}} \left( k \right) = |\mathcal{R}_{c}|^{2}$. The dimensionless power spectrum is 
\begin{equation}
{\mathcal{P}}_{\mathcal{R}} = \frac{k^3}{2 {\pi}^2} |\mathcal{R}_{c}|^{2}, \\
\end{equation}
\noindent and the spectral index is defined as
\begin{equation}
{n}_{s} \equiv 1 + \frac{d \ln {\mathcal{P}}_{\mathcal{R}}}{d \ln {k}_{\text{hc}}}, \\
\end{equation}
\noindent where ${k}_{\text{hc}} = a \left( {t}_{\text{hc}} \right) H \left( {t}_{\text{hc}} \right)$ represents the pivot scale at the first horizon crossing ${t}_{\text{hc}}$, which is related to the cosmic time $t$ by
\begin{equation}
\frac{d \ln{k}}{dt} = \frac{d \left(aH \right)}{d t} = \frac{\dot{a}}{a} + \frac{\dot{H}}{H} = H \left( 1 + \frac{\dot{H}}{H^2} \right) = \left( 1 - {\epsilon} \right) H. \\
\end{equation}
Using the transfer function, we can relate the power spectra of adiabatic and entropic perturbations at time ${t}_{\text{hc}}$ to its value at some later time $t > {t}_{\text{hc}}$ with the corresponding pivot scale $k$ as
\begin{equation}
\begin{split}
{\mathcal{P}}_{\mathcal{R}} \left( k \right) =& \; {\mathcal{P}}_{\mathcal{R}} \left( {k}_{\text{hc}} \right) \left[ 1 + {T}^{2}_{\mathcal{RS}} \left( {t}_{\text{hc}}, t \right) \right], \\
{\mathcal{P}}_{\mathcal{S}} \left( k \right) =& \; {\mathcal{P}}_{\mathcal{R}} \left( {k}_{\text{hc}} \right) {T}^{2}_{\mathcal{SS}} \left( {t}_{\text{hc}}, t \right), \\
\end{split}
\end{equation}
\noindent The transfer functions are given by
\begin{equation}
\begin{split}
\frac{1}{H} \frac{{\partial} {T}_{\mathcal{RS}}}{{\partial} {t}_{\text{hc}}} &=\; - {\alpha} - {\beta} {T}_{\mathcal{RS}}, \\
\frac{1}{H} \frac{{\partial} {T}_{\mathcal{SS}}}{{\partial} {t}_{\text{hc}}} &=\; - {\beta} {T}_{\mathcal{SS}}. \\
\end{split}
\end{equation}
\noindent In term of the number of e-folding $N$, the above differential equation becomes
\begin{equation}
\begin{split}
\frac{{\partial} {T}_{\mathcal{RS}}}{{\partial} {N}_{\text{hc}}} &=\; - {\alpha} - {\beta} {T}_{\mathcal{RS}}, \\
\frac{{\partial} {T}_{\mathcal{SS}}}{{\partial} {N}_{\text{hc}}} &=\; - {\beta} {T}_{\mathcal{SS}}. \\
\end{split}
\end{equation}
\noindent The spectral index for the power spectrum of the adiabatic fluctuations becomes
\begin{equation}
{n}_{s} \simeq {n}_{s} \left( {t}_{\text{hc}} \right) + \frac{1}{H} \left( \frac{{\partial} {T}_{\mathcal{RS}}}{{\partial} {t}_{\text{hc}}} \right) \sin{2 \Delta}, \\
\end{equation}
\noindent where
\begin{equation}
{n}_{s} \left( {t}_{\text{hc}} \right) = 1 - 6 {\epsilon} \left( {t}_{\text{hc}} \right) + 2 {\eta}_{{\sigma}{\sigma}} \left( {t}_{\text{hc}} \right), \\
\end{equation}
\noindent and the trigonometric functions for $T_{\mathcal{RS}}$ are defined as
\begin{equation}
\label{Trigo Delta}
\begin{split}
\sin{\Delta} \equiv& \; \frac{1}{\sqrt{1 + T^{2}_{\mathcal{RS}}}}, \\ 
\cos{\Delta} \equiv& \frac{{T}_{\mathcal{RS}}}{\sqrt{1 + {T}^{2}_{\mathcal{RS}}}}, \\
\tan{\Delta} \equiv& \; \frac{1}{T_{\mathcal{RS}}}. \\
\end{split}
\end{equation}
\noindent The iso-curvature fraction is given by
\begin{equation}
{\beta}_{\text{iso}} \equiv \frac{{\mathcal{P}}_{\mathcal{S}}}{{\mathcal{P}}_{\mathcal{R}} + {\mathcal{P}}_{\mathcal{S}}} = \frac{T^{2}_{\mathcal{SS}}}{1 + T^{2}_{\mathcal{SS}} + T^{2}_{\mathcal{RS}}}, \\
\end{equation}
\noindent which can be used for compared with the recent observables in Planck collaboration. Also, the tensor-to-scalar ratio is given by
\begin{equation}
r \simeq \frac{16 {\epsilon}}{1 + T^{2}_{\mathcal{RS}}}. \\
\end{equation}

\section{Numerical Calculations and Results}
\label{Numerical Calculations and Results}
\noindent Now, we carry out numerical calculations. For simplicity, we consider the minimal coupling case that we take the kinetic terms as canonical in both Jordan and Einstein frames, which means $\tilde{\mathcal{G}}_{IJ} = \mathcal{G}_{IJ}={\delta}_{IJ}$. It then follows that $f \left( {\phi}^{I} \right) = \frac{1}{2}{M}^{2}_{\text{pl}}$. The potential function becomes

\begin{equation}
V({\phi}_{1}, {\phi}_{2}) = \Lambda ^4 R^4 \left(1-\cos \left(\frac{\phi _1}{f_1}+\frac{\phi _2}{f_2}\right)\right)+\Lambda ^4 \left(1-\cos \left(\frac{\phi_1}{g_1}+\frac{\phi_2}{g_2}\right)\right). \\
\end{equation}
\noindent Also, for verifying the parameter sets with Planck observation, we should recall the following
\begin{equation}
{n}_{s} \simeq 1 - 6 {\epsilon} \left( {N}_{\text{hc}} \right) + 2 {\eta}_{{\sigma}{\sigma}} \left( {N}_{\text{hc}} \right) - \left({\alpha} \left( {N}_{\text{hc}} \right) + {\beta} \left( {N}_{\text{hc}} \right) \cot{\Delta} \right) \sin{2 \Delta}, \\
\end{equation}
\begin{equation}
r \simeq 16 {\epsilon}\left( {N}_{\text{hc}} \right) \sin^{2}{\Delta}, \\
\end{equation}
\begin{equation}
{\beta}_{\text{iso}} = \frac{T^{2}_{\mathcal{SS}} \left( {N}_{\text{hc}}, {N} \right)}{1 + T^{2}_{\mathcal{SS}} \left( {N}_{\text{hc}}, {N} \right) + T^{2}_{\mathcal{RS}} \left( {N}_{\text{hc}}, {N} \right)}, \\
\end{equation}
\begin{equation}
\cos{\Delta} \equiv \frac{{T}_{\mathcal{RS}}}{\sqrt{1 + {T}^{2}_{\mathcal{RS}}}}. \\
\end{equation}

\noindent The parameters for the rest of the paper correspond to the above potential function. In our numerical calculations, we take the common parameters as shown in Table \ref{tab: Common parameters for numerical calculations}. 

\begin{figure}[h]
\centering
\includegraphics[width=70mm, height=70mm]{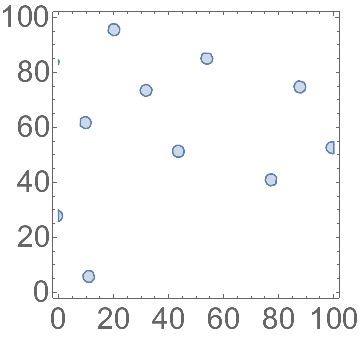} 
\includegraphics[width=70mm, height=70mm]{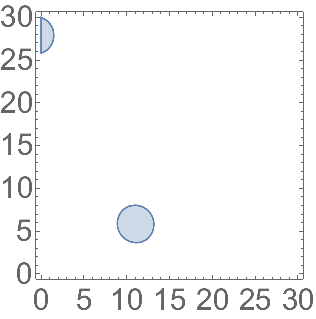} 
\caption{Feasible initial ${\phi}_{1}$ and ${\phi}_{2}$ under the constraints of Planck 2018 observation as shown in Table \ref{table:Planck data 2018 slow roll potential parameters and spectral indices} are shown on the left. The $x$ axis and $y$ axis of both graphs correspond to the initial values of ${\phi}_{1}$ and ${\phi}_{2}$ in units of ${M}_{\text{pl}}$ respectively. For further investigation, we focus on the circle $9 {M}_{\text{pl}} \leq {\phi}_{1} \leq 13 {M}_{\text{pl}}$ and $3 {M}_{\text{pl}} \leq {\phi}_{2} \leq 8 {M}_{\text{pl}}$ as shown on the right. }
\label{fig: Initial phi1 and phi2}
\end{figure}

\begin{figure}[h]
\centering
\includegraphics[width=70mm, height=50mm]{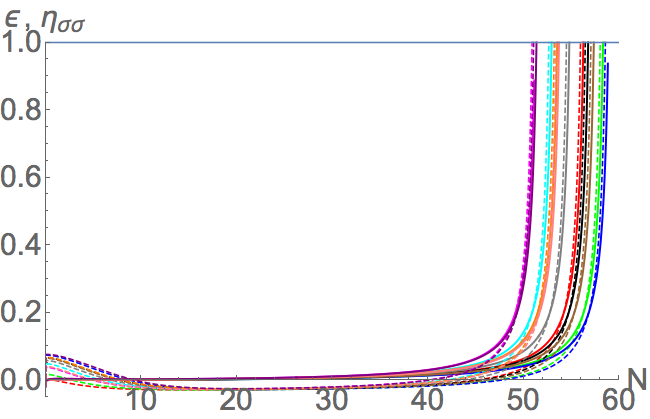} 
\includegraphics[width=70mm, height=50mm]{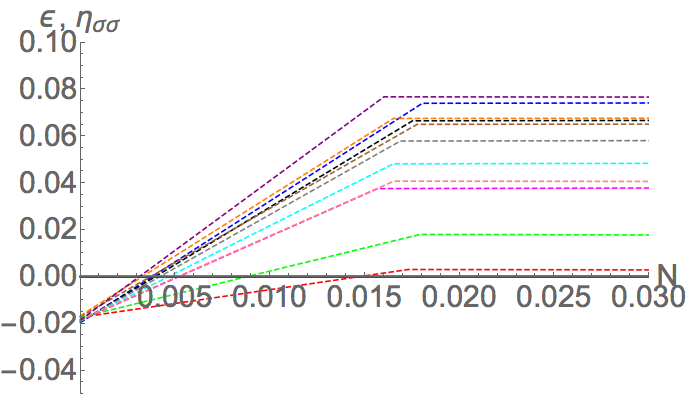} 
\caption{Evolutions of ${\epsilon}$ (solid lines) and ${\eta}$ (dashed lines) are shown on the left. The initial evolutions are shown on the right. Since all the parameter sets have small $\epsilon$ value at scales of about $O \left( 10^{-9} \right)$, they overlap around the $x$ axis. Different colours correspond to the parameter sets listed in Table \ref{tab: Initial phi1 and phi2 and the corresponding results}. }
\label{fig: Epsilon and Eta}
\end{figure}

\begin{table}[h]
\begin{center}
\begin{tabular}{ |c|c|c|c|c|c|c|c|c|c|c| }
\hline
${N}_{\text{hc}}$ & ${\phi}'_{1} \left( {N}_{\text{hc}} \right)$ & ${\phi}'_{2} \left( {N}_{\text{hc}} \right)$ & $R$ & ${\Lambda}/{M}_{\text{pl}}$ & ${f}_{1} / {M}_{\text{pl}}$ & ${f}_{2} / {M}_{\text{pl}}$ & ${g}_{1}/ {M}_{\text{pl}}$ & ${g}_{2}/ {M}_{\text{pl}}$ \\
\hline
$0$ & $1 \times 10^{-5}$ & $1 \times 10^{-5}$ & $0.73$ & $0.005$ & $4.5$ & $8.5$ & $8.5$ & $-4.5$ \\
\hline
\end{tabular}
\end{center}
\caption{Common parameters for numerical calculations}
\label{tab: Common parameters for numerical calculations}
\end{table}
\noindent Based on the constraints in Planck 2018 observation tabled before, we make a region plot for feasible initial values of ${\phi}_{1}$ and ${\phi}_{2}$ as shown in Figure \ref{fig: Initial phi1 and phi2}. One can see the shaded region on the region plot, which corresponds to the feasible initial field values. Quantitatively, the tolerance allowance of less fine tuning is $4 {M}_{\text{pl}}$ for ${\phi}_{1}$ and $5 {M}_{\text{pl}}$ for ${\phi}_{2}$. For demonstration, we take the following initial values of ${\phi}_{1}$ and ${\phi}_{2}$, denoted by ${\phi}_{1 \text{ini}}$ and ${\phi}_{2\text{ini}}$ respectively as shown in Table \ref{tab: Initial phi1 and phi2 and the corresponding results}. Different colours correspond to the plots in the rest of this paper. The number of e-folding at the end of inflation is evaluated at the minimum value of $N$ such that either one of the slow-roll parameters ${\epsilon}$ or ${\eta}$ becomes $1$. As a check, one may see the evolutions of ${\epsilon}$ and ${\eta}_{{\sigma}{\sigma}}$ in Figure \ref{fig: Epsilon and Eta}.  \\

\begin{table}[h]
\begin{center}
\begin{tabular}{ |c|c|c|c|c|c|c|c|c|c|c| }
\hline
$\text{Color}$ & ${\phi}_{1 \text{ini}} / {M}_{\text{pl}}$ & ${\phi}_{2 \text{ini}} / {M}_{\text{pl}}$ & ${N}_{\text{end}}$ & ${\beta}_{\text{iso}}$ & $\cos{\Delta} \left( \times 10^{-4} \right)$ \\
\hline
$\text{Red}$ & $9.9$ & $5$ & $56.4$ & $1.27562 \times 10^{-34}$ & $2.58297$ \\
\hline
$\text{Green}$ & $10$ & $5$ & $58.4$ & $3.56105 \times 10^{-35}$ & $2.32159$ \\
\hline
$\text{Blue}$ & $11.9$ & $7$ & $58.8$ & $8.91444 \times 10^{-36}$ & $1.21171$ \\
\hline
$\text{Black}$ & $12$ & $7$ & $57.1$ & $5.31969 \times 10^{-35}$ & $1.2716$ \\
\hline
$\text{Gray}$ & $12.1$ & $7$ & $55$ & $1.29562 \times 10^{-33}$ & $1.33897$ \\
\hline
$\text{Cyan}$ & $12.2$ & $7$ & $53.3$ & $1.05749 \times 10^{-40}$ & $1.41531$ \\
\hline
$\text{Magenta}$ & $12.3$ & $7$ & $51.1$ & $2.15349 \times 10^{-36}$ & $1.50251$ \\
\hline
$\text{Brown}$ & $12.5$ & $6$ & $57.7$ & $9.84201 \times 10^{-46}$ & $2.04699$ \\
\hline
$\text{Orange}$ & $12.7$ & $6$ & $53.8$ & $1.02946 \times 10^{-36}$ & $1.72768$ \\
\hline
$\text{Pink}$ & $9.5$ & $5.5$ & $53.9$ & $1.79031 \times 10^{-36}$ & $3.02397$ \\
\hline
$\text{Purple}$ & $9.2$ & $5.8$ & $51.5$ & $6.92661 \times 10^{-48}$ & $1.30517$ \\
\hline
\end{tabular}
\end{center}
\caption{Initial ${\phi}_{1}$ and ${\phi}_{2}$ and the corresponding ${\beta}_{\text{iso}}$ and $\cos{\Delta}$ evaluated from ${N}_{\text{hc}}$ and ${N}_{\text{end}}$. Colours correspond to the curves/lines in all the graphs of this paper. }
\label{tab: Initial phi1 and phi2 and the corresponding results}
\end{table}

\vspace{3mm}

\begin{figure}[h]
\centering
\includegraphics[width=150mm, height=90mm]{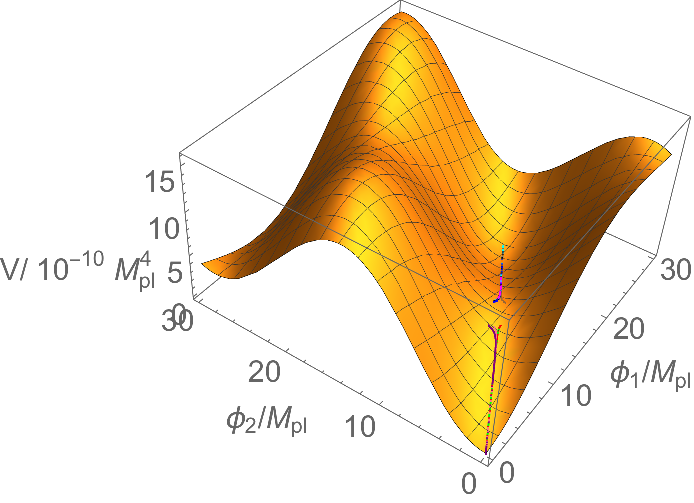} 
\caption{Natural inflation potential and field evolutions. The colours of the curves correspond to the parameters listed in Table \ref{tab: Initial phi1 and phi2 and the corresponding results}.}
\label{fig: Natural Potential}
\end{figure}

\begin{figure}[h]
\centering
\includegraphics[width=150mm, height=100mm]{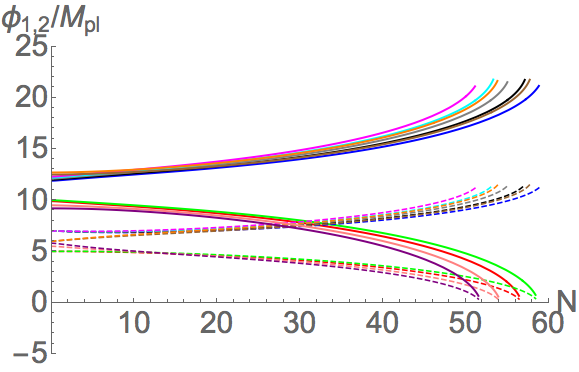} 
\caption{Field evolution as the number of e-foldings runs from $0$ to some values between $50$ and $60$. The solid lines correspond to the evolution of ${\phi}_{1}$ while the dashed lines correspond to that of ${\phi}_{2}$. The colours refer to the parameters listed in Table \ref{tab: Initial phi1 and phi2 and the corresponding results}.}
\label{fig: Field Values of Evolution}
\end{figure}

\noindent Based on the above parameters, we plotted the evolution of fields along the potential well as shown in Figure \ref{fig: Natural Potential}. Starting from the initial values listed above, inflation carries out as the curves roll down correspondingly and reach the troughs at the end of inflation. We can see that red and green lines roll down to the trough located at the point $\left( {\phi}_{1}, {\phi}_{2} \right) = \left( 0,0 \right)$, while other colour lines roll down to another trough located at the point $\left( {\phi}_{1}, {\phi}_{2} \right) = \left( 21,11 \right)$. This can be further confirmed by taking a look at Figure \ref{fig: Field Values of Evolution}, which shows the field evolutions as the number of e-foldings run. After that, the fields oscillate about the troughs and reheating occurs. 

\begin{figure}[h]
\includegraphics[width=150mm, height=100mm]{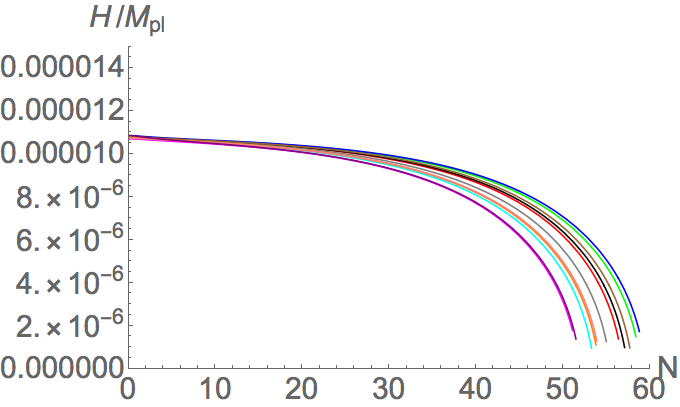} 
\caption{Evolution of Hubble parameter against the number of e-foldings. Starting from about $1.1 \times 10^{-5}$, Hubble parameter initially decreases slowly and decreases significantly after about $30$ e-folding. }
\label{fig: Hubble Parameter Evolution}
\end{figure}

\begin{figure}[h]
\includegraphics[width=70mm, height=50mm]{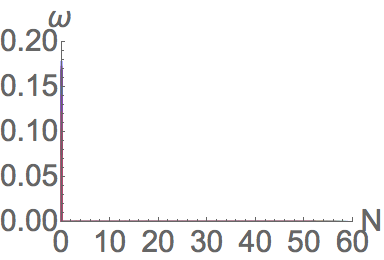} 
\includegraphics[width=70mm, height=50mm]{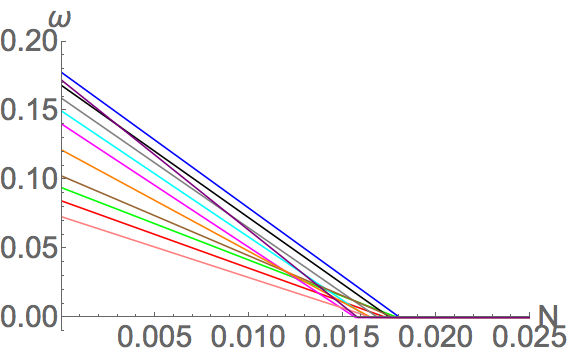} 
\caption{Turn rate against the number of e-foldings from $0$ to $60$ are shown on the left. The turn rate drops very quickly to $0$. On the right, one may see the initial evolutions of $\omega$, say, from $0$ e-folding to $0.04$ e-folding. }
\label{fig: Turn Rate Evolution}
\end{figure}

\noindent Next, we plot the evolutions of Hubble parameter under various parameter sets and initial conditions as shown in Figure \ref{fig: Hubble Parameter Evolution}. Starting from about $1.1 \times 10^{-5}$, Hubble parameter initially decreases slowly and decreases significantly after about $30$ e-foldings. This matches with the observation constraint $\frac{H}{{M}_{\text{pl}}}<2.7 \times 10^{-5}$ \cite{Planck 2018 X} and confirms our usual assumption that $H$ remains nearly constant during slow-roll approximation and ${\epsilon}=- \frac{\dot{H}}{H^2}$ gradually tends to $1$ as inflation ends. Furthermore, we plot the turn rate as shown in Figure \ref{fig: Turn Rate Evolution}. Basically, the turn rate initially drops quickly from about $0.07 \sim 0.16$ to $0$. This implies that the inflation curve initially turns a little bit and moves along a straight line down to a trough without changing its direction, which matches with the inflation curves shown in Figure \ref{fig: Natural Potential}.  

\vspace{3mm}

\begin{figure}[h]
\includegraphics[width=75mm, height=50mm]{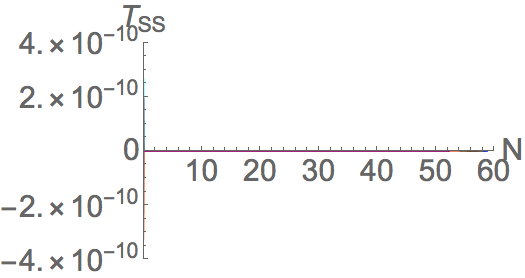} 
\includegraphics[width=75mm, height=50mm]{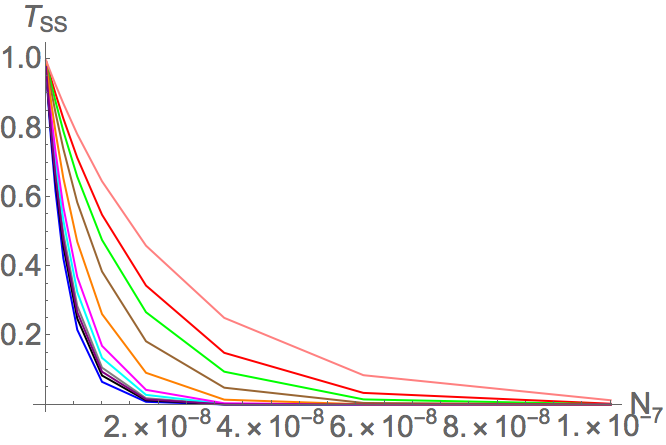} 
\caption{Evolutions of ${T}_{\mathcal{SS}}$ against the number of e-foldings are shown on the left. The graph on the left shows the evolutions for the whole inflation, while the graph on the right shows the initial evolutions. }
\label{fig: TSS Evolution}
\end{figure}

\begin{figure}[h]
\includegraphics[width=75mm, height=50mm]{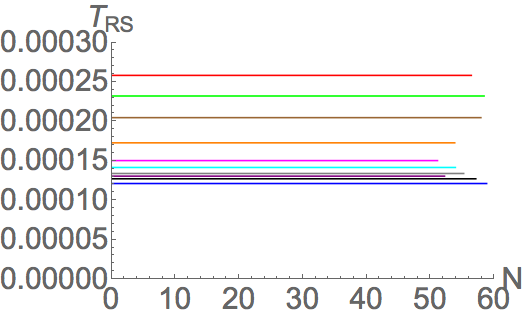} 
\includegraphics[width=75mm, height=50mm]{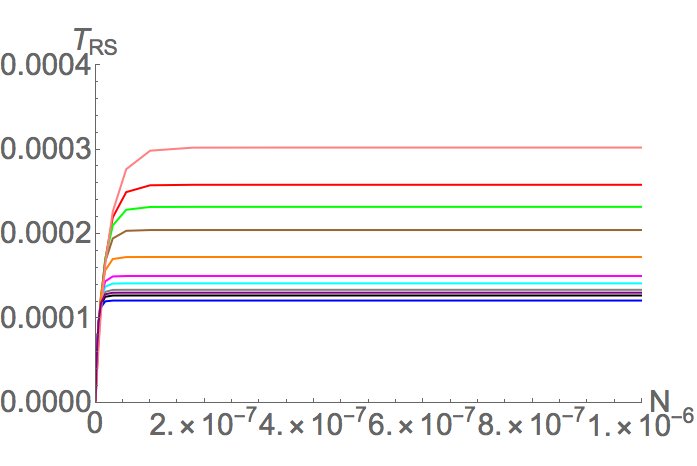} 
\caption{Evolutions of ${T}_{\mathcal{RS}}$ against the number of e-foldings are shown on the left. The graph on the left shows the evolutions for the whole inflation, while the graph on the right shows the initial evolutions. }
\label{fig: TRS Evolution}
\end{figure}

\begin{figure}[h]
\includegraphics[width=75mm, height=50mm]{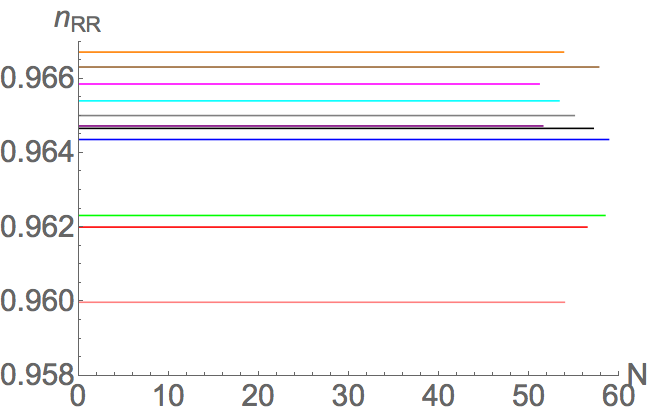} 
\includegraphics[width=75mm, height=50mm]{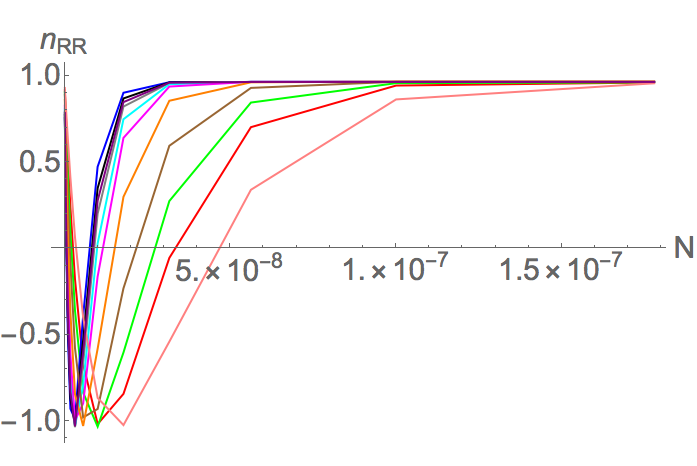} 
\caption{Evolutions of ${n}_{\mathcal{RR}}$ against the number of e-foldings. The graph on the left shows the evolutions for the whole inflation, while the graph on the right shows the initial evolutions. }
\label{fig: nRR Evolution}
\end{figure}

\noindent In addition, the evolutions of ${T}_{\mathcal{SS}}$ and ${T}_{\mathcal{RS}}$ against the number of e-foldings are plotted in Figure \ref{fig: TSS Evolution} and Figure \ref{fig: TRS Evolution}. For ${T}_{\mathcal{SS}}$, starting from $1$, it decreases rapidly to reach nearly zero for the rest of the inflation process after $N=O \left( 10^{-7} \right)$. Meanwhile, for ${T}_{\mathcal{RS}}$, starting from $0$, it rises significantly at a decreasing rate and remains stable after $N=O \left( 10^{-7} \right)$. Finally, graphs of spectral indexes against the number of e-foldings are plotted in Figure \ref{fig: nRR Evolution}. The evolutions initially have a sudden drop and they remain level between $0.96$ and $0.967$ after $N=O \left( {10}^{-7} \right)$. Since the tensor-to-scalar ratio is given by $r \simeq \frac{16 {\epsilon} \left( {N}_{\text{hc}} \right)}{1 + {T}_{\mathcal{RS}} \left( {N}_{\text{hc}}, N \right)^{2}}$ and ${\epsilon}=O \left( 10^{-9} \right)$ in our model, it satisfies the latest Planck constraints \cite{Planck 2018 X}. Furthermore, ${\beta}_{\text{iso}}$ and $\cos{\Delta}$ evaluated from ${N}_{\text{hc}}$ to ${N}_{\text{end}}$ are listed on Table \ref{tab: Initial phi1 and phi2 and the corresponding results}. Basically, we can see that ${\beta}_{\text{iso}}$ and $\cos{\Delta}$ have scales of at least $O \left( 10^{-30} \right)$ and $O \left( 10^{-4} \right)$ respectively, which satisfy the latest observation constraint ${\beta}_{\text{iso}} < 9.5 \times 10^{-4}$ and $-0.05 \leq \cos{\Delta} \leq 0.05$.

\begin{figure}[h]
\includegraphics[width=75mm, height=50mm]{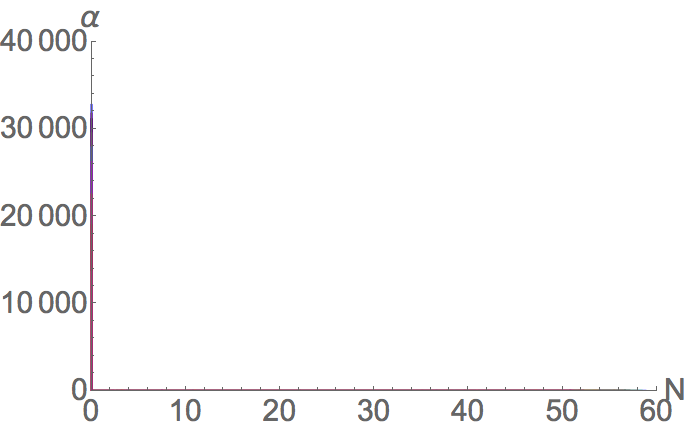} 
\includegraphics[width=75mm, height=50mm]{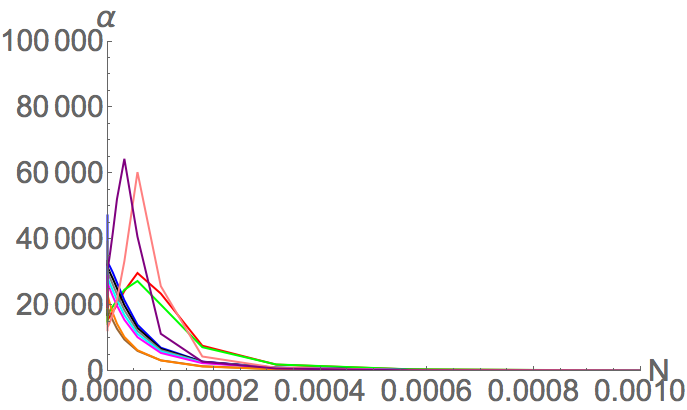} 
\caption{Evolutions of ${\alpha}$ against the number of e-foldings. The graph on the left shows the evolutions for the whole inflation, while the graph on the right shows the initial evolutions. }
\label{fig: Alpha Evolution}
\end{figure}

\begin{figure}[h]
\includegraphics[width=75mm, height=50mm]{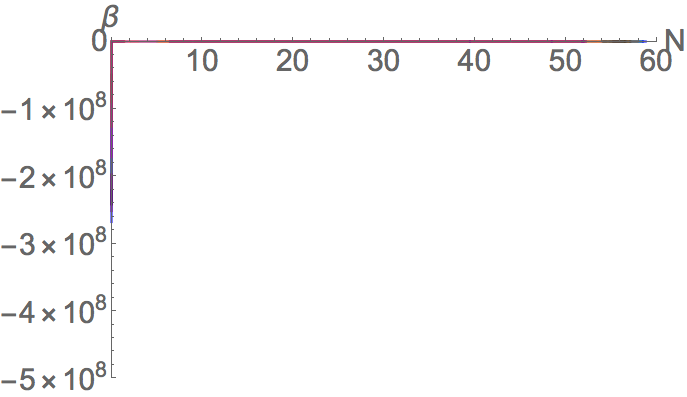} 
\includegraphics[width=75mm, height=50mm]{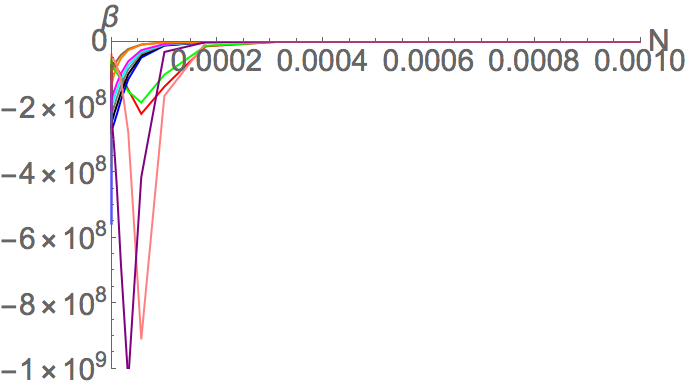} 
\caption{Evolutions of ${\beta}$ against the number of e-foldings. The graph on the left shows the evolutions for the whole inflation, while the graph on the right shows the initial evolutions. }
\label{fig: Beta Evolution}
\end{figure}

\section{Discussion}
\label{Discussion}
\noindent In this section, we try to explain some results listed above and make comments on them. First of all, about the smallness of $\epsilon$, this is because the initial velocities of field values are taken to be small. Now that we take the initial velocities at a scale of $O \left( 10^{-5} \right)$, ${\epsilon}$ initially has a scale of about $O \left( 10^{-9} \right)$. For example, if we replace the initial velocities at a scale of $O \left( 10^{-3} \right)$, then ${\epsilon}$ initially has a scale of about $O \left( 10^{-5} \right)$ instead, which also satisfies the latest Planck observation \cite{Planck 2018 X}, and vice versa. 

\vspace{3mm}

\noindent Next, one may see the evolutions of ${T}_{\mathcal{SS}}$ and ${T}_{\mathcal{RS}}$ change significantly at the beginning of inflation and remain constant during the inflation. To know the reasons, we should take a look at the evolutions of ${\alpha} \left( N \right)$ and ${\beta} \left( N \right)$ as shown in Figure \ref{fig: Alpha Evolution} and \ref{fig: Beta Evolution}. For the evolution of ${\alpha} \left( N \right)$, we can see that all curves decay rapidly and go to zero after $N=4 \times 10^{-4}$. This makes ${T}_{\mathcal{RS}}$ initially surge and then remain level. The reason why ${\alpha} \left( N \right)$ have such a evolution is because the smallness of Hubble parameter at the beginning of inflation, which have a scale of $O \left( 10^{-5} \right)$ as shown in Figure \ref{fig: Hubble Parameter Evolution}, contributes to a large scale, and the turn rate ${\omega}$ drops to zero after $N=0.02$ as shown in Figure \ref{fig: Turn Rate Evolution}. Similarly, for the counterpart of  ${\beta} \left( N \right)$, we can see that all curves rise at a very fast rate starting from $-10^{9}$ to reach zero after $N=3 \times 10^{-4}$. This makes ${T}_{\mathcal{SS}}$ decays rapidly and remains nearly zero after $N=10^{-7}$. Due to these two reasons, they make the spectral indexes suffer a sudden drop to about $-1$ and return to the range between $0.96$ and $0.967$ as shown in Figure \ref{fig: nRR Evolution}. These parameter sets describe a scenario that the entropic perturbation mainly contribute to the adiabatic perturbation instead of self-preserving. Non-perturbative effects of supergravity and superstring theory can trigger inflation.

\section{Conclusions and Future Work}
\noindent In summary, we adopt the double field natural inflation model motivated by the non-perturbative effects of supergravity and superstring theory to do the slow roll analysis. We can see that under the constraints given by Planck observation there is freedom ($4 {M}_{\text{pl}}$ for ${\phi}_{1}$ and $5 {M}_{\text{pl}}$ for ${\phi}_{2}$) to choose the initial field values for inflation. This is in favour of our conjecture that quantum fluctuation causes the uncertainty of initial field values. Not only do they satisfy the Planck observation, but it also gives us some interesting physical ideas like how entropic perturbation contributes to adiabatic perturbation as supporting the universe to inflate. For the direction of further investigation, non-Gaussianity and double natural inflation with minimal and non-minimal coupling will be interesting. Double field natural inflation with non-canonical kinetic terms is also an interesting direction for further exploration. We also expect that more precise observations can be made so that stricter constraints can be attained to help us find the correct dynamics.

\section{Acknowledgments}
I thank Prof. Hiroyuki ABE very much for suggestion of my research, Dr. Hajime OTSUKA and Mr. Shuntaro AOKI for useful discussions related to field stabilisations and supergravity potential evaluations. I also want to express my sincere gratitude to Prof. Keiichi MAEDA for his opinions towards cosmological inflation analysis.

\end{document}